\documentstyle[12pt,epsfig]{article}
\newcommand{\be}{\begin{equation}}
\newcommand{\ee}{\end{equation}}
\newcommand{\ba}{\begin{eqnarray}}
\newcommand{\ea}{\end{eqnarray}}

\newcommand{\ga}{\gamma_5}
\newcommand{\dg}{^{\dagger}}
\newcommand{\la}{\lambda}
\newcommand{\re}[1]{(\ref{#1})}
\newcommand{\na}[1]{\nabla_{#1}}
\newcommand{\Id}{\mbox{1\hspace{-0.98mm}l}}   
\newcommand{\mb}[1]{\quad\mbox{ #1 }\quad}
\newcommand{\e}{\mbox{e}}
\newcommand{\cao}{{\cal O}} 
\newcommand{\F}{{\cal F}} 
\newcommand{\pt}{\partial} 
\newcommand{\di}{\mbox{d}\,}

\newcommand{\G}{\Gamma} 
\newcommand{\bG}{\bar{\Gamma}} 
\newcommand{\df}{[\mbox{d}\bar{\psi}\mbox{d}\psi]}
\newcommand{\dfp}{[\mbox{d}\bar{\psi}'\mbox{d}\psi']}
\newcommand{\f}{{\mbox{\tiny f}}}

\begin{document}
\renewcommand{\baselinestretch}{1.1} \small\normalsize

\begin{flushleft}
   May 1999 \hfill {\sc HUB-EP}-99/21
\end{flushleft}

\vspace*{.9cm}

\begin{center}

{\Large \bf Dirac operator normality and chiral properties
                     \\on the lattice}

\vspace*{0.8cm}

{\bf Werner Kerler}

\vspace*{0.3cm}

{\sl Institut f\"ur Physik, Humboldt-Universit\"at, D-10115 Berlin, 
Germany}
\hspace{3.6mm}

\end{center}

\vspace*{2.5cm}

\begin{abstract}
Normality in connection with $\gamma_5$-hermiticity determines the basic 
chiral properties and rules. The Ginsparg-Wilson (GW) relation is 
one of the allowed constraints on the spectrum. Interrelations between
features of the spectrum, the sum rule for chiral differences of real modes
and contributions to the Ward identity are pointed out. The alternative 
chiral transformation of L\"uscher gives the same Ward identity as the usual 
one, in the global and in the local case. Imposing normality on a general 
function of the hermitean Wilson-Dirac (HWD) operator, inevitably leads at the
same time to the Neuberger operator and to the GW relation. In this context 
also the case with zero eigenvalues of the HWD operator is handled. The 
eigenvalue flows of the HWD operator obey a differential equation the 
characteristic features of the solutions of which are discussed. 
\end{abstract}

\vspace*{.2cm}

\newpage

\section{Introduction and overview}

\hspace{3mm}
Recently considerable progress in the description of chiral fermions has been
initiated by works of Neuberger \cite{ne98} from the point of view of the
overlap formalism \cite{na93} and of Hasenfratz et al.~\cite{ha98} in the 
context of fixed point actions. This has also revived considerations of
the Ginsparg-Wilson (GW) relation \cite{gi82} which in both cases turned out 
to be satisfied. On the basis of the GW relation L\"uscher \cite{lu98} 
proposed an alternative chiral transformation providing chiral symmetry at the 
classical level. The finite form of this transformation has been given
by Chiu \cite{ch98}. 

Neuberger \cite{ne98}, in particular, was able to derive an explicit form of 
the massless Dirac operator based on the hermitean Wilson-Dirac (HWD) operator,
which also plays a major r\^ole in the overlap formalism \cite{na93}. These
developments have then given rise to several numerical studies of 
eigenvalue flows of the HWD operator with the mass parameter
\cite{na97}, relying on the fact that this operator has well defined 
spectral properties.

Despite the many publications which followed the mentioned works, there
are clearly still many questions open. In the present paper
we address a number of them which are related to a more precise 
understanding and to basic properties of these new developments. 

We start from the observation that the Dirac operator must be normal in order 
that reliable conclusions become possible. This follows from two theorems
of the spectral theory of operators in unitary spaces. The first one of them
says that normality is necessary and sufficient in order that the eigenvectors 
form a complete sytem. Apart from the fact that otherwise very little is known 
on spectral properties, this implies that without normality there are 
necessarily defects which can hardly be tolerated in making predictions.

The second one of these theorems more specifically concerns chiral properties.
It states that normality guarantees that an eigenvector of the operator is
at same time an eigenvector of the adjoint operator. This together with 
$\ga$-hermiticity (i.e.~hermiticity of the operator multiplied by $\ga$) 
is exactly what provides the basis for chiral behaviors. 

We point out that, given normality and $\ga$-hermiticity, one already obtains
the basic rules and properties. Apart from the general structure of the
eigenvector system, the fundamental sum rule for chiral differences of real 
modes emerges. The relations between the operator and such modes, and in 
particular those to the index, as needed in various contexts, immediately 
follow.

A further consequence of normality is that the operator decomposes into 
commuting operators related to the real part and to the imaginary part
of its eigenvalues. This allows to study possible constraints on the spectrum. 
In particular, restricting it to a one-dimensional set, we find that, in
addition to zero, the curve must meet the real axis at least at one further
point, in order that the sum rule mentioned above allows a nonzero index. From 
this point of view the general nature of the GW relation becomes clear.
It is just one of such constraints which satisfies the requirement 
of a further real eigenvalue in a minimal way.

The symmetry at the classical level provided by the alternative chiral 
transformation \cite{lu98,ch98} makes things similar to what one is accustomed 
to in continuum theory, however, at the price of complications due to the 
action dependence of this transformation. At the quantum level, where there is 
anyway no such symmetry, the question arises, what the precise difference 
to the usual chiral transformation is. 

In order to have a basis for a general comparison of transformations, we 
derive the Ward identities in an appropriately general way. It is
seen that by the normality of the operator the global chiral
transformation leads just to the sum rule for chiral differences of real modes. 
The correspondence of the contributions to terms familiar
in continuum theory is pointed out. The results of the local chiral 
transformation are similarly identified.

It is shown that the alternative chiral transformation leads exactly to the 
same result as the usual chiral one, even without assuming the GW relation. 
Imposing the GW relation has the only effect to specialize the results
to that case with only two real eigenvalues. We also give the local version 
of the alternative transformation. The resulting Ward identity is again 
seen to agree with that of the usual local chiral one.

The operator of Neuberger is the only explicit form of a massless Dirac
operator on the lattice presently known. In order to get more insight into the
possibilities of the construction of such operators it appears desirable to
have also a derivation of it which does not rely on the overlap formalism. 
Further, there is the somewhat unsatisfactory point that zero eigenvalues 
of the HWD operator so far had to be excluded. 

In our derivation of the indicated operator the requirement of normality
is central. To avoid doublers and at the same time to deal with well defined
mathematical properties, at present the only possibility is to start from the
HWD operator. Therefore we consider a general function of this operator
and impose the necessary conditions on it. Doing this it turns out that 
the requirement of normality is an extremely strong one, leading at the same 
time to the Neuberger form of the Dirac operator and to the GW constraint
on the spectrum. In addition the inclusion of zeros of the HWD operator 
gets nonstraightforward. Nevertheless, also for this problem a way out is 
found.

For the explicit Dirac operator the relations of its eigenvectors to that 
of the HWD operator become transparent. The apparent importance of the HWD 
operator suggests to study its eigenvalue flows with the mass parameter also
analytically. We show that they satisfy a differential equation and give a 
complete overview of the characteristic properties of its solutions. 

Section 2 is devoted to the basic chiral properties. In Section 3 possible
constraints on the spectrum are discussed. Section 4 contains the general 
derivation of Ward identities. In Section 5 the results for particular 
transformations are analyzed. Section 6 gives the systematic construction of 
a normal operator. In Section 7 the spectral flows of the HWD
operator are investigated.

\section{Basic chiral operator properties} \setcounter{equation}{0}

\hspace{3mm}
We require $D$ to be normal
\be
[D,D\dg]=0 
\label{nor}
\ee
and $\ga$-hermitean
\be
D\dg = \ga D \ga  \quad .
\label{Ddg}
\ee
Because of \re{nor} the solutions $f_k$ of the eigenequation
\be
D f_k = \la_k f_k
\label{eg}
\ee
form a complete orthonormal set, on the basis of which general conclusions
become possible. By \re{nor} simultaneous eigenvectors of $D$ and $D\dg$ 
exist. In the present context this has the important consequence that one 
also has 
\be
D\dg f_k = \la_k^* f_k  
\label{egdg}
\ee
which together with \re{Ddg} leads to 
\be
D\ga f_k = \la_k^*\ga f_k \quad .
\label{eg5}
\ee
The comparison of \re{eg} multiplied by $\ga$ with \re{eg5} then gives
\be
[\ga,D] f_k = 0 \quad \mbox{ if } \quad \la_k \mbox{ real } ,
\label{egc}
\ee
which tells that in the subspace of real eigenvalues of $D$ one can 
introduce simultaneous eigenvectors of $D$ and of $\ga$, i.e.~ones with 
chirality.

Multiplying \re{eg} from the left by $f_l\dg \ga$ and its adjoint 
$f_l\dg D\dg = f_l\dg \la_l^*$ from the right by $\ga f_k$ one obtains the
relation
\be
f_l\dg \ga f_k = 0 \quad \mbox{ for } \quad \la_l^* \ne \la_k \quad ,
\label{nela}
\ee
which actually reflects the orthogonality of eigenvectors related to 
different eigenvalues of $D$. In view of \re{egc} and of the comparison
of \re{eg5} with \re{eg}, respectively, it is 
convenient to introduce in more detail
\be
f_k = \left\{\begin{array}{l@{\quad\mbox{for}\quad}l@{\quad\mbox{with}\quad}l}
f_k^{(5)}& \mbox{Im}\la_k=0&                     \ga f_k^{(5)}=c_k f_k^{(5)}\\ 
f_k^{(1)}& \mbox{Im}\la_k > 0& \ga f_k^{(1)}=f_k^{(2)}\\
f_k^{(2)}& \mbox{Im}\la_k < 0& \ga f_k^{(2)}=f_k^{(1)}
  \end{array} \right. 
\label{sD}
\ee
where for the chirality $c_r$ possible values are $+1$ and $-1$. Obviously
eigenvectors related to complex eigenvalues always come in pairs, while 
those related to real ones need not to do so. Of course, how many of each 
type occur depends on the particular $D$, however, by \re{nor} and 
\re{Ddg}, the structure of the eigenvector system is in any case the one 
described. 

For the numbers of modes related to a real eigenvalue $\la$ of $D$ 
\be
N_{\pm}(\la) = \sum_{k\atop \la_k=\la \,\mbox{\tiny  real }} 
                        \frac{1\pm c_k}{2}            
\label{Nr}
\ee
from \re{nela}, \re{sD} and Tr$(\ga)=0$ one obtains the sum rule for the
chiral differences of real modes 
\be
   \sum_{\la \mbox{ \tiny  real }} \Big(N_+(\la) - N_-(\la)\Big) = 0 \quad.
\label{res}
\ee
It in particular implies that the difference for eigenvalue zero 
$N_-(0) - N_+(0)$, the index of $D$, can only be nonvanishing if a 
corresponding difference from nonzero eigenvalues exists.  

 From \re{nela}, \re{sD} and \re{Nr} one also readily gets the useful relations
\be
\varepsilon \mbox{Tr}((D+\varepsilon)^{-1}\ga) \rightarrow N_+(0) - N_-(0)
       \quad \mbox{ for } \quad \varepsilon \rightarrow 0
\label{re0}
\ee
\be
   \mbox{Tr}((D+\varepsilon)^{-1}\ga D) \rightarrow 
   \sum_{\la\ne 0 \mbox{ \tiny  real }} \Big(N_+(\la) - N_-(\la)\Big)
       \quad \mbox{ for } \quad \varepsilon \rightarrow 0
\label{re1}
\ee
\be
    \mbox{Tr}(\ga D)=
   \sum_{\la\ne 0 \mbox{ \tiny  real }} \la\,\Big(N_+(\la) - N_-(\la)\Big)
\label{rel}
\ee
between $D$ and $N_+(\la) - N_-(\la)$. It will be discussed later that 
\re{re0} and \re{re1} are related to the index term and the topological 
charge ($F\tilde{F}$) term, respectively, of continuum theory. They obviously 
add up to the sum rule \re{res}. If only one nonzero value for real eigenvalues
occurs, one can also use \re{rel} instead of \re{re1} to relate $D$ to the
corresponding chiral difference.

\section{Constraints on location of spectrum}  \setcounter{equation}{0}

\hspace{3mm}
In order to study possible constraints on the spectrum of $D$ we use
the decomposition
\be
D = u + i v \quad \mbox{ with } \quad 
u= u\dg= \frac{1}{2} (D+D\dg) \quad , \quad v= v\dg = \frac{1}{2i} (D-D\dg)
\label{Duv}
\ee
for which \re{nor}, the normality of $D$, implies 
\be
[u,v]=0     \quad .
\label{uvc}
\ee 
By \re{Ddg}, the $\ga$-hermiticity of $D$, it follows that
\re{Duv} is at the same time the decomposition into the parts commuting and 
anticommuting with $\ga$ and that $[\ga,u]=0$ and $\{\ga,v\}=0$ hold.
According to \re{uvc} $u$, $v$ and $D$ have simultaneous eigenvectors. The 
real eigenvalues of $u$ and $v$ are simply the real and imaginary parts,
respectively, of those of $D$. Therefore we can specify the location of 
the spectrum of $D$ by constraining $u$ and $v$. 

A particular simplification arises if the spectrum is located on a 
one-dimensional set. This can be realized by imposing the condition 
\be
\F(u,v)=0
\label{F0}
\ee
 with a suitably chosen function $\F(u,v)$ which, as a function 
of commuting hermitean operators, is well defined. In the notation of \re{eg}, 
it satisfies the eigenequation 
$\F(u,v) f_k = \F(\mbox{Re}\la_k,\mbox{Im}\la_k) f_k$ and has the  
spectral representation 
$\F(u,v)=\sum_k  \F(\mbox{Re}\la_k,\mbox{Im}\la_k) f_k f_k\dg$.
Because we wish to allow for the eigenvalue $0$ of $D$, the function $\F$ 
considered as a function of real numbers should have 
the property $\F(0,0)=0$. Since we in addition want that the index 
of $D$ can be nonzero, according to \re{res} also at least 
one real eigenvalue different from zero must be possible. Therefore in addition
one has to require  $\F(\beta,0)=0$ for at least one real $\beta\ne0$. Thus
the function must have the particular properties 
\be
\F(0,0)=0 \mb{and} \F(\beta,0)=0 \mb{for some} \beta \ne 0
\label{FF}
\ee
in order that the curve specified by \re{F0} meets the real axis at 
zero and at least at one further point.

A simple possibility which satisfies condition \re{FF} is a circle through
zero with center on the real axis
\be
\F(u,v) = (u-\rho)^2 + v^2 - \rho^2 = 0 \quad.
\label{gwF}
\ee
Using \re{Duv} and \re{uvc} one can write \re{gwF} as $\rho(D+D\dg)=D\dg D$
which by \re{Ddg} is seen to be just the GW relation
\be
\{\ga,D\}= \rho^{-1} D \ga D   \quad .
\label{gw}
\ee
In contrast to the original form \cite{gi82}, however, \re{gw} does not 
involve a further operator sandwiched in addition to $\ga$ into its 
r.h.s.~because this would spoil the normality of $D$. Therefore here only 
a real constant $\rho^{-1}$ remains. Obviously \re{gwF} meets the requirement 
\re{FF} in a minimal way, admitting only $2\rho$ in addition to $0$. The sum 
rule \re{res} then simplifies to two terms. This conforms with the observation 
of Chiu \cite{ch98a} that in the case of the GW relation $\{\ga,D\}= D \ga D$ 
the chiral differences obtained at $0$ and at $2$ add up to zero. 
 
By \re{FF} the choice $\F=u$, corresponding to $\{\ga,D\}=0$, is excluded 
because there is only one real eigenvalue (on the finite lattice with $D$ 
being bounded a second one can also not occur at $\infty$). This choice may, 
however, be approached in the continuum limit. In fact, considering the 
stereographic projection of the circle \re{gwF} on the sphere of complex 
numbers it is seen that for increasing radius it approaches the circle 
through $\infty$ on this sphere which corresponds to the imaginary axis 
in the plane. With $\rho\sim 1/a$ for decreasing lattice spacing $a$ the 
envisaged approach indeed occurs. This suggests that in the continuum the 
sum rule for chiral differences could possibly be satisfied by eigenvalues 
at $0$ and at $\infty$. Of course, the subtleties of the respective limit 
remain to be investigated.

The form of the constraint on the spectrum depends on the particular 
properties of the operator $D$ considered. The Neuberger operator is tied to 
the GW constraint. For other constructions in any case also \re{FF} is to be 
required. In addition, also the appropriate behavior in the limit should be 
guaranteed.

\section{General form of the Ward identity} \setcounter{equation}{0}

\hspace{3mm}
In dealing with Ward identities it should be remembered that the expectation
values 
\be
\langle \cao \rangle = \frac{\int[\mbox{d}U]\df \e^{-S_U-S_\f}\,\cao}
                            {\int[\mbox{d}U]\df \e^{-S_U-S_\f}}
\mb{with} S_\f=\bar{\psi}M\psi 
\label{exp}
\ee
involve integrals $\int\df \e^{-S_\f} = \det M $ and
\be 
\int\df \e^{-S_\f}\psi_{j_1}\bar{\psi}_{k_1}\ldots\psi_{j_s}\bar{\psi}_{k_s}= 
\sum_{l_1\ldots l_s}\epsilon^{k_1\ldots k_s}_{l_1\ldots l_s}
            M^{-1}_{j_1 l_1}\ldots M^{-1}_{j_s l_s} \det M 
\label{int}
\ee
(where $\epsilon^{k_1\ldots k_s}_{l_1\ldots l_s}=+1,\,-1$ or $0$ if 
${k_1\ldots k_s}$ is an even, odd or no permutation of ${l_1\ldots l_s}$).
Thus, in order that the expectation values \re{exp} are properly defined, 
$M^{-1}$ must exist and $\det M$ be nonzero. Therefore, to be able to proceed 
in the presence of zero modes of $D$ one has to put $M=D+\varepsilon$ and 
let $\varepsilon$ go to zero in the final result. 

Fermionic Ward identities arise from  the condition that 
$\int \df \e^{-S_\f} \cao $ must not change under the transformation  
\be
\psi'= \exp(i\eta\G)\psi \mb{,} \bar{\psi}'= \bar{\psi} 
       \exp(i\eta\bar{\G}) \quad ,
\label{tC}
\ee
where $\eta$ is a parameter. This means that one gets the identity 
\be
\frac{\di}{\di\eta} \int\dfp \e^{-S_\f'} \cao' \Big|_{\eta=0} = 0 
\label{w0}
\ee
with three contributions, one from the derivative of the integration
measure, one from that of the action and one from that of $\cao$.
For the measure contribution within 
$\dfp=\df\Big(\det\exp(i\eta\bar{\G)}\det\exp(i\eta\G)\Big)^{-1}$
one obtains
\be
\frac{\di(\det\exp(i\eta\bar{\G})}{\di\eta}\Big|_{\eta=0} = i\mbox{Tr }\bar{\G}
\mb{,}\frac{\di(\det \exp(i\eta\G)}{\di\eta}\Big|_{\eta=0} = i\mbox{Tr }\G\quad.
\label{ms}
\ee
With respect to the derivative of $\cao$ we note that for products 
${\cal P}$ of $\psi$ and $\bar{\psi}$ fields, since the Grassmann-even 
combinations $\psi_j\frac{\pt}{\pt\psi_k}$ and 
$\bar{\psi}_j\frac{\pt}{\pt\bar{\psi}_k}$ can be readily 
shifted to the appropriate place, one can relate 
\be
\frac{\di {\cal P}'}{\di\eta} \Big|_{\eta=0} = 
i\sum_l \Big( (\G\psi)_l\frac{\pt {\cal P}}{\pt \psi_l} +
(\bar{\psi}\bG)_l\frac{\pt{\cal P}}{\pt \bar{\psi}_l} \Big) \quad.
\label{os}
\ee
The fermionic part of $\cao$ in general is made up of such products and of 
sums thereof. More specifically it can even be considered to be made up of 
products of equal numbers of $\psi$ and $\bar{\psi}$ fields because only 
such products contribute to the intergrals; for the same reason $\cao$ can 
also be considered to be Grassmann-even. Thus in any case \re{os} applies 
to $\cao$ and \re{w0} becomes 
\be
i\int\df \e^{-S_\f}\Big(-\mbox{Tr}(\bar{\G}+\G)\cao 
-\bar{\psi}(\bar{\G}M+M\G)\psi\cao + 
\bar{\psi}\bG\frac{\pt \cao}{\pt \bar{\psi}} -
\frac{\pt \cao}{\pt \psi}\G\psi \Big) = 0 \quad .
\label{w1}
\ee 
This generalizes the relation \cite{bo85} which, with suitable choices of
$\cao$, is used in many applications to the case where the integration 
measure is not invariant.

In studies of the singlet axial vector current and its relation to anomaly,
index and topological charge it usually suffices to consider the case $\cao=1$,
as is e.g.~also done in Ref.~\cite{ha98}. To keep things general we avoid
this here, integrating out the $\bar{\psi}$ and $\psi$ fields in the second 
term of \re{w1} without specifying $\cao$. Then at the same time that term 
gets on equal footing with the first one, as is desirable for a convenient 
comparison of transformations. To integrate out the indicated fields we use 
the identity
\ba
0&=&\frac{1}{2}\int\df\Bigg(
\Big(\frac{\pt}{\pt\psi}M^{-1}\Big)_j(\e^{-S_\f}\psi_k\cao) +
\Big(M^{-1}\frac{\pt}{\pt\bar{\psi}}\Big)_k(\e^{-S_\f}
\bar{\psi}_j\cao)\Bigg)
\nonumber \\&=&
\int\df\e^{-S_\f}\Bigg(\bar{\psi}_j\psi_k\cao + 
M^{-1}_{kj}\cao  
+\frac{1}{2}\Big(\frac{\pt \cao}{\pt \psi} M^{-1}\Big)_j \psi_k 
-\frac{1}{2}\bar{\psi}_j\Big(M^{-1} \frac{\pt \cao}{\pt\bar{\psi}}
  \Big)_k \Bigg) 
\label{Ir}
\ea
which relies on the fact that $\int\df(\pt/\pt\psi_l)G=0$ and
$\int\df(\pt/\pt\bar{\psi}_l)G=0$ for any function $G$. Then \re{w1} becomes
\be
iW \int\df \e^{-S_\f}\cao + \frac{i}{2}\int\df\e^{-S_\f}\Big(
\frac{\pt \cao}{\pt \psi}M^{-1}R\psi + 
\bar{\psi}RM^{-1}\frac{\pt \cao}{\pt\bar{\psi}} \Big) = 0 
\label{w2}
\ee
where $R = \bar{\G}M-M\G$ and
\be 
 W = \mbox{Tr}\Big(-\bar{\G}-\G + M^{-1}(\bar{\G}M+M\G)\Big) \quad .
\label{w}
\ee
To evaluate the terms with derivatives of $\cao$ in \re{w2} further we
remember that the fermionic part of $\cao$ can be considered to be made up 
of products of type 
${\cal P}= \psi_{j_1}\bar{\psi}_{k_1}\ldots\psi_{j_s}\bar{\psi}_{k_s}$
for which by \re{int} we find
\ba
-\int\df\e^{-S_\f} \frac{\pt {\cal P}}{\pt \psi}M^{-1}R\psi =
+\int\df\e^{-S_\f} \bar{\psi}RM^{-1}\frac{\pt {\cal P}}{\pt\bar{\psi}}  =
\nonumber\\
s\sum_{l_1\ldots l_s}\epsilon^{k_1\ldots k_s}_{l_1\ldots l_s}
M^{-1}_{j_1 l_1}\ldots M^{-1}_{j_s-1 l_s-1}
(M^{-1}RM^{-1})_{j_s l_s} \det M \quad .
\ea
This shows that the terms in \re{w2} with derivatives of $\cao$ cancel
and we remain with
\be
iW \int\df \e^{-S_\f}\cao =0 \quad ,
\ee
which inserted into \re{exp} gives the Ward identity $\langle W\cao\rangle=0$ 
or, if desired, also the one in a background gauge field 
$\,W\langle \cao\rangle_U=0$.

\section{Results for particular transformations} \setcounter{equation}{0}

\hspace{3mm}
For the global chiral transformation, which in terms of \re{tC} is given by
\be
\G=\bar{\G}=\ga \quad ,
\label{t5g}
\ee
the measure contribution $-\mbox{Tr}(\bG+\G)$ vanishes and one obtains 
\be
W= \mbox{Tr}(M^{-1}\{\ga,M\}) 
\label{WM}
\ee 
or inserting $M=D+\varepsilon$ 
\be
W = \mbox{Tr}\Big((D+\varepsilon)^{-1}\{\ga,D\}\Big)+
2\varepsilon\mbox{Tr}\Big((D+\varepsilon)^{-1}\ga\Big)  \quad .
\label{gwa2}
\ee
The first term in \re{gwa2} by \re{re1} is seen to become the sum over 
$2(N_+(\la) - N_-(\la))$ for nonvanishing real $\la$ and the 
second one by \re{re0} the difference $2(N_+(0) - N_-(0))$. By \re{gwa2} 
they add up to 
\be
W \rightarrow 2\sum_{\la \mbox{ \tiny  real }} \Big(N_+(\la) - N_-(\la)\Big)
\mb{for} \varepsilon \rightarrow 0 \quad .
\label{was}
\ee
Thus we obviously arrive just at the sum rule for chiral differences of 
real modes \re{res}. Analoguous to the features known in continuum 
theory for quite some time \cite{br77}, the first term in \re{gwa2} 
is the the topological charge ($F\tilde{F}$) term while the second one is 
the index term. The latter by \re{re0} is obvious. For the first term 
the limit has been established long ago \cite{ke81} for the Wilson-Dirac 
operator and recently \cite{ad98} also for the Neuberger operator (for
which the use of $\mbox{Tr}(\ga D)$ by \re{re1} and \re{rel} with only one
nonzero $\la$ is equivalent to using 
$\mbox{Tr}\Big((D+\varepsilon)^{-1}\{\ga,D\}\Big)$).

Next we consider the alternative transformation \cite{lu98,ch98} for which
in our notation
\be
\G=\ga(1-(2\rho)^{-1}M) \mb{,} \bar{\G}=(1-(2\rho)^{-1}M)\ga \quad .
\label{tag}
\ee
With \re{tag} one now gets $-\mbox{Tr}(\bG+\G)=+\rho^{-1}\mbox{Tr}(\ga M)$ 
for the measure contribution and $\mbox{Tr}\big(M^{-1}(\bG M+M\G))=
\mbox{Tr}(M^{-1}\{\ga,M\})-\rho^{-1}\mbox{Tr}(\ga M)$ for the action 
contribution. Obviously the extra term of the latter cancels the measure 
term so that again the result \re{WM} is obtained, and notably, even
without assuming the GW relation. 

If with the alternative transformation in addition the GW relation \re{gw} 
is imposed, inserting $M=D+\varepsilon$, for the action contribution
one gets $\mbox{Tr}\big((M^{-1}(\bG M+M\G))= 
2\varepsilon(1+(2\rho)^{-1}\varepsilon)\mbox{Tr}((D+\varepsilon)^{-1}\ga)$
which by \re{re0} becomes $2(N_+(0) - N_-(0))$. For the measure contribution 
one has $-\mbox{Tr}(\bG+\G)\rightarrow \rho^{-1}\mbox{Tr}(\ga D)$ which
by \re{rel} equals 
$\sum_{\la\ne 0 \mbox{ \footnotesize  real }} \la(N_+(\la) - N_-(\la))$.
In the GW case with only $\la=2\rho$ this becomes 
$2(N_+(2\rho) - N_-(2\rho))$. Taking both contributions\footnote{In 
  Ref.~\cite{lu98} the action contribution is missing. The evaluation of the 
  measure contribution there is actually a use of the identity 
  $0=\mbox{Tr}\ga= \varepsilon\mbox{Tr}((D+\varepsilon)^{-1}\ga)
 +\mbox{Tr}((D+\varepsilon)^{-1}\ga D)$ which by inserting the
  GW relation \re{gw} becomes 
  $0=\varepsilon(1+\varepsilon(2\rho)^{-1})\mbox{Tr}((D+\varepsilon)^{-1}\ga) 
  +(2\rho)^{-1}\mbox{Tr}(\ga D)$. }
together it is obvious that one gets the same results as before now 
specialized to the case where only $0$ and $2\rho$ occur for real eigenvalues.

The local chiral transformation in the present context is conveniently 
introduced by 
\be
\G=\bar{\G}=\ga\hat{e}(n) \mb{with} 
	 \Big(\hat{e}(n)\Big)_{n''n'}=\delta_{n''n}\delta_{nn'} 
\label{t5}
\ee
for which \re{w} gives 
\be
W= \mbox{Tr}\Big((M^{-1}\{\ga\hat{e}(n),M\}\Big) \quad .
\label{WMe}
\ee
By inserting the decomposition 
$M=\frac{1}{2}(M-\ga M\ga)+\frac{1}{2}(M+\ga M\ga)$ 
(into parts anticommuting and commuting with $\ga$) and also $M=D+\varepsilon$ 
into $\{\ga\hat{e}(n),M\}$ this becomes
\be
W = \frac{1}{2}\mbox{Tr}\Big((M^{-1}[\hat{e}(n),[\ga,D]\,]\Big)  +
\frac{1}{2}\mbox{Tr}\Big((M^{-1}\{\hat{e}(n),\{\ga,D\}\}\Big)   +
         2\varepsilon\mbox{Tr}\Big((M^{-1}\ga\hat{e}(n)\Big) \, .
\label{loc}
\ee
The first term in \re{loc} is seen to vanish upon summation over $n$ and 
accordingly corresponds to the divergence of the singlet axial vector current. 
Summation over $n$ in the rest, responsible for current nonconservation, 
leads to the results of the global transformation. The second term in \re{loc}
in the limit gives the $F\tilde{F}$-density of continuum theory 
\cite{ke81,ad98}. The third term in \re{loc} is the local version of the index 
contribution. To visualize things in terms of current expressions one should 
remember that the $M^{-1}$ factors in \re{loc} correspond to the integrated 
out $\bar{\psi}$ and $\psi$ fields. 

We note that the local transformation related to the alternative chiral
transformion \re{tag} can also be introduced. It is given by 
\be
\G=\ga\hat{e}(n)(1-(2\rho)^{-1}M) \mb{,} \bar{\G}
                           =(1-(2\rho)^{-1}M)\ga\hat{e}(n) \quad .
\label{ta}
\ee
The calculation of $W$ with this transformation again leads to \re{WMe} so 
that it becomes obvious that also in the local case nothing new is obtained.

Of course, also other transformations could straightforwardly be considered
along the present lines. For the nonsinglet chiral one with flavor 
operator $T_l$ one has  $\G=\bar{\G}=\ga T_l$ in the global and 
$\G=\bar{\G}=\ga T_l\hat{e}(n)$ in the local case, with current conservation 
resulting from $\mbox{Tr}\,T_l=0$. Conserved vector currents are related
to $\G=-\bar{\G}=T_l\hat{e}(n)$ in the nonsinglet case and to 
$\G=-\bar{\G}=\hat{e}(n)$ in the singlet case.

\section{Derivation of normal operator} \setcounter{equation}{0}

\hspace{3mm}
The Wilson-Dirac operator $X/a$ (with hermitean $\gamma$-matrices in 
4-dimensional euclidean space and $0<r\le 1$) is given by
\be
X = \frac{r}{2} \sum_{\mu} \na{\mu}\dg\na{\mu} + m
     + \frac{1}{2} \sum_{\mu} \gamma_{\mu}(\na{\mu}-\na{\mu}\dg) 
\label{DW}
\ee
where $(\na{\mu})_{n'n} = \delta_{n'n} - U_{\mu n} \delta_{n',n+\hat{\mu}}$ 
(which
implies $\na{\mu}\dg\na{\mu}=\na{\mu}\na{\mu}\dg=\na{\mu}+\na{\mu}\dg$). 
For the operator $X$ one has $\ga$-hermiticity,
\be
X\dg = \ga X \ga \quad , 
\label{Xdg}
\ee
however, in the presence of a gauge field (with 
$[\na{\mu},\na{\nu}]\ne 0$ and $[\na{\mu}\dg,\na{\nu}]\ne 0$ for 
$\mu \ne \nu$ and thus $[X\dg,X]\ne 0$) $X$ is not normal. 

To derive a normal and $\ga$-hermitean operator $D$ one needs to start from 
\be
H=\ga X
\label{H}
\ee
which, being hermitean, in contrast to $X$ has well defined spectral 
properties. The strategy then is, instead of $X=\ga H$, to consider 
\be
D=\ga E(H)+C
\label{YE}
\ee
with some general function $E(H)$ and some constant $C$, and to determine
those quantities by imposing the necessary conditions.  
Requiring $\ga$-hermiticity \re{Ddg} of $D$ it follows that 
$E(H)$ must be hermitean and that $C$ must be real.  Since with 
\be
H \phi_l = \alpha_l \phi_l 
\label{egH}
\ee
(where $\alpha_l$ is real and the $\phi_l$ form a complete orthonormal 
set) one has the representation 
$ E(H) = \sum_l E(\alpha_l) \phi_l \phi_l\dg$, hermiticity of $E(H)$ simply
means that $E(\alpha)$ considered as a function of a real parameter 
$\alpha$ must be a real function. 
 
 From the requirement of normality \re{nor} of $D$ we obtain the condition 
\be
[\ga,E(H)^2]=0   \quad .
\label{5E2}
\ee
To satisfy this condition is the central point. Wishing to get a general
solution, one must require $E(H)^2$ to be independent of $H$. Though being
inevitable, this is clearly quite drastic. It means that $E(H)^2$ should be 
a multiple of the identity 
\be
E(H)^2=\rho^2\Id \quad ,
\label{const}
\ee
or that $E(\alpha)=\pm\rho$ with some constant $\rho$ which, without 
restricting generality, we can take to be positive. In order to keep the 
properties of $E(H)$ as close as possible to those of $H$ we further require 
$E(\alpha)$ to be nondecreasing and odd. This fixes the signs and we end up 
with 
\be
E(\alpha) = \rho \,\epsilon(\alpha) 
\label{Ea}
\ee
where $\epsilon(\alpha)=\pm 1$ for $\alpha{>\atop <} 0$. 
If all $\alpha_l\ne 0$ this is already the solution and $E(H)/\rho$ is 
just the function $H/\sqrt{H^2}$ of Neuberger \cite{ne98}.   
If $\alpha_l=0$ occur we have to specify $\epsilon(0)$. It would be tempting
to take the value zero for this,\footnote{
  Which, by the way, would lead to 
  $\F(u,v)=\big((u-\rho)^2 + v^2\big)\big((u-\rho)^2+v^2-\rho^2\big) = 0$
  instead of \re{gwF}.}
however, because of \re{const}, i.e.~the necessity to keep the procedure 
independent of $H$, this is definitely not possible. Thus one has to
choose either $+1$ or $-1$ for $\epsilon(0)$, and one must decide for one
of them since no $H$-independent criterion for selection appears available. 
The oddness of the function, which then is violated at $\alpha=0$, can be 
recovered by doing independent calculations for each of the two choices and 
taking the mean of the final results. In Section 7 we will see that in terms 
of counting eigenvalue flows this procedure has a natural equivalent.

To fix the constant $C$ in \re{YE} we note that, because $\rho^{-1}\ga E(H)$ 
is unitary, the spectrum of $\ga E(H)$ is on the circle with radius $\rho$
and center at zero. Thus to get the appropriate spectrum of $D$ we put 
$C=\rho$ and get
\be
D=\rho\,(1+\ga\epsilon(H)) \quad . 
\label{DV}
\ee 
It appears important to emphasize that, by the necessity to satisfy \re{const},
one cannot escape simultaneously arriving at the Ginsparg-Wilson constraint 
\re{gwF} on the spectrum and at the Neuberger form \re{DV} of the Dirac 
operator. In addition \re{const} unavoidably produces the somewhat delicate 
situation with the choice of $\epsilon(0)$.

To complete the derivation the occurring parameters are to be fixed. The
continuum limit in the case $U=\Id$ with the representation 
$(\na{\mu})_{pp} = 1-\e^{-ip_{\mu}a}$ indicates that masslessness requires
$m<0$ and one gets $\rho=|m|/a$. Further, it is also known that to avoid
effects of doublers on the finite lattice one needs $m>-2r$. A choice with 
major analytical simplifications is $-m=r=1$. It should be noted that $X$ 
enters \re{DV} only up to a positive constant factor, so that, for example,
using $X/a$ instead of $X$ would not change anything.

Since real eigenvalues occur only at $0$ and at $2\rho$, the
sum rule \re{res} for the operator \re{DV} reduces to 
$N_+(0) - N_-(0) + N_+(2\rho) - N_-(2\rho)= 0$. Similarly
\re{rel} becomes $\mbox{Tr}(\ga D)= 2\rho(N_+(2\rho) - N_-(2\rho))$.
Combining these two relations one has 
$N_-(0) - N_+(0)=(2\rho)^{-1}\mbox{Tr}(\ga D)$ and inserting the
particular form \re{DV} of $D$ into this one gets
\be
N_-(0) - N_+(0)=\frac{1}{2}\mbox{Tr}(\epsilon(H))
\label{ind}
\ee
for its index. Also the eigenvectors of $D$ and of $H$ can now be related in 
detail. For this we note that \re{eg} in terms of $\ga\epsilon(H)$ becomes 
$\ga\epsilon(H) f_k = (\la_k/\rho-1) f_k$, so that parametrizing complex
eigenvalues as $\la_k=\rho(1+e^{i\varphi_k})$ with $0<\varphi_k<\pi$ and
remembering \re{sD} we get the eigenequations of $\epsilon(H)$ 
\ba
\epsilon(H) f_k^{(5)} & = & \mp c_k f_k^{(5)} \mb{for} \la_k=
    \left\{\begin{array}{r@{}l}
    0\\ 2\rho \end{array} \right. \nonumber\\
\epsilon(H) f_k^{(\pm)} & = & \pm f_k^{(\pm)}
\mb{with} f_k^{(\pm)} = 
\frac{1}{\sqrt{2}} (\e^{-i\varphi_k/2}f_k^{(1)} \pm 
                    \e^{i\varphi_k/2} f_k^{(2)}) \quad .
\label{ze}
\ea
These equations are to be compared with 
\be 
\epsilon(H) \phi_l = \epsilon(\alpha_l) \phi_l 
\label{ege}
\ee
which results from \re{egH}. Obviously the vectors in \re{ze} are linear 
combinations $\tilde{\phi}^{\pm}_k=\sum_l b_{kl}^{(\pm)}\phi_l^{(\pm)}$ 
where $\phi_l^{(\pm)}= \phi_l$ for $\epsilon(\alpha_l)=\pm 1$. By the
properties of $D$ derived here and the general structure obtained in
Section 2 it is guaranteed that task to find the coefficients  
$b_{kl}^{(\pm)}$ has a solution.

\section{Relations for spectral flows} \setcounter{equation}{0}

\hspace{3mm}
The studies of the flows of eigenvalues of $H$ with $m$ can be justified
on the basis of \re{ind} which in terms of $N_+^H$ and $N_-^H$, the numbers 
of positive and negative eigenvalues of $H$, in the absence of 
eigenvalues zero of $H$ reads 
\be
N_-(0) - N_+(0)= \frac{1}{2} (N_+^H - N_-^H) \quad .
\label{reNN}
\ee
The crossing of zero of an eigenvalue which occurs at some $m$ is connected 
to a change of the difference of the numbers of positive and negative 
eigenvalues by $+2$ or $-2$, respectively, depending on the direction of the 
crossing. Therefore the net number of crossings is related to the index of $D$.

To include also zero eigenvalues of $H$ in these considerations, one has to
note that in the very moment of crossing a positive (negative) eigenvalue
has disappeared, however, a negative (positive) one has not yet appeared.
Still using \re{reNN} then agrees with the notion that the index in that 
moment has only changed by $\frac{1}{2}$. With this understanding \re{reNN} 
is no longer equivalent to \re{ind} in which $\epsilon(0)=0$ is forbidden 
by \re{const}. However, the analogue of the procedure described in Section 6 
(of working with the mean of the choices $\epsilon(0)=+1$ and $\epsilon(0)=-1$),
in the case of counting flows is seen to lead to the same result as the 
counting at the crossing point mentioned above. Thus the latter appears 
valid and natural.

To investigate properties of the flows of eigenvalues of $H$ analytically we 
first derive some relations. Multiplying \re{egH} by $\phi_l\dg\ga$ one gets 
$\phi_l\dg\ga H \phi_l = \alpha_l \phi_l\dg\ga \phi_l$ and summing this
and its hermitian conjugate one has
$\phi_l\dg \{\ga,H\}\phi_l = 2\alpha_l \phi_l\dg\ga \phi_l$. From this
by inserting \re{H} with \re{DW} one obtains
\be
 \alpha_l\, \phi_l\dg\ga \phi_l = g_l(m) + m  \mb{with}    g_l(m) =
\frac{r}{2} \sum_{\mu} ||\na{\mu}\phi_l||^2  
\label{gm}
\ee
(where  $0\le g_l(m)\le 8r$ since 
$||\na{\mu}\phi_l||\le||(\na{\mu}-\Id)\phi_l||+||\phi_l||=2$). 
Next, abbreviating $(\di \alpha_l)/(\di m)$ by $\dot{\alpha}_l$, we note
that
\be
\frac{\di (\phi_l\dg H \phi_l)}{\di m}=\phi_l\dg \dot{H} \phi_l +
\dot{\phi}_l\dg H \phi_l+\phi_l\dg H \dot{\phi}_l =
\phi_l\dg\ga \phi_l+\alpha_l \frac{\di (\phi_l\dg \phi_l)}{\di m}  
\ee
which means that we have
\be
\dot{\alpha}_l=\phi_l\dg\ga \phi_l \quad .
\label{pm}
\ee
Combining \re{gm} and \re{pm} we get the differential equation
\be
\dot{\alpha}_l \alpha_l=g_l(m) + m 
\label{dif}
\ee
which can be readily integrated to give 
\be
\alpha_l^2(m)=\alpha_l^2(m_b)+2\int^m_{m_b}\di m'(m'+g_l(m'))
	     = \alpha_l^2 (m_b)+m^2-m_b^2+2\int^m_{m_b}\di m'g_l(m') 
\label{dint}
\ee
in which particular solutions are determined by the choice of 
$\alpha_l^2 (m_b)$.

To get an overview of the set of solutions we note that because of 
$H\rightarrow m\ga$ for $m \rightarrow \pm \infty$ one gets 
$g_l(m)\rightarrow 0$ for $m \rightarrow \pm\infty$. Therefore, since $g_l(m)$ 
is nonnegative, the equation $m+g_l(m)=0$ has at least one solution
$m_0\le 0$. If this is the only one we choose $m_b=m_0$. Because then
$\int^m_{m_0}\di m'(m'+g_l(m'))\ge 0$ for all $m$ it becomes obvious that
we can freely choose $\alpha_l^2(m_0)\ge 0$. In this way we get all solutions
allowed by \re{pm}, which requires $\dot{\alpha_l}$ to be finite (the 
forbidden solutions can be conveniently seen by choosing $\alpha_l^2(m_b)= 0$ 
and $m_b\ne m_0$).  The extension to the general case, where one has to deal 
with $2z+1$ solutions of $m+g(m)=0$ with 
$m_{2z}\le m_{2z-1}\le\ldots\le m_1\le m_0$, is straightforward. Then among 
the $m_y$ with even $y$ one has to equate that to $m_b$ which leads to 
the lowest value of
$\int^{m_{y}}_{\tilde{m}}\di m'(m'+g_l(m')$ for some fixed $\tilde{m}$ 
(or one of those in case of degeneracy). 

We thus have a complete specification of the solutions. Clearly all solutions 
obtained show the asymptotic behaviors 
$\alpha_l^2(m) \rightarrow m^2$ for $ m^2 \rightarrow \infty$.
It is seen that the points $m_y$ with even $y$ determine the characteristic 
features. If $\alpha_l^2(m_y)>0$ there is a minimum of the solution 
$+\sqrt{\alpha_l^2(m)}$ and a maximum of the solution $-\sqrt{\alpha_l^2(m)}$  
at that point. If $\alpha_l^2(m_y)=0$ then $+\sqrt{\alpha_l^2(m)}$
coming from above continues as $-\sqrt{\alpha_l^2(m)}$ below the zero,
and analogously $-\sqrt{\alpha_l^2(m)}$ from above as $+\sqrt{\alpha_l^2(m)}$
below, i.e.~one gets two solutions which cross zero at that point.
For the square of the derivative at the crossing point
(using $\dot{\alpha}_l^2=(\dot{\alpha}_l \alpha_l)^2/\alpha_l^2$ and \re{dif}) 
one obtains 
\be
\dot{\alpha}_l^2(m) \rightarrow 1+\dot{g}_l(m) \mb{for} m \rightarrow m_y
                          \mb{and} \alpha_l^2(m_y)=0 
\ee 
which shows that $\dot{g}_l(m)$ in general will cause deviations from the
chiral value 1. The solutions of the differential equations describe the 
possibilities for flows which occur. Which values of $\alpha_l^2 (m_b)$
and which signs of $\pm\sqrt{\alpha_l^2(m)}$ are selected and what the 
detailed properties of the function $g_l(m)$ are depends on the eigenequation 
\re{egH} which in the present context no longer appears directly.

\section*{Acknowledgement}

\hspace{3mm}
I wish to thank Michael M\"uller-Preussker and his group for the warm
hospitality extended to me.


\newpage

\begin{thebibliography}{9}

\bibitem{ne98}  H. Neuberger, 
              Phys. Lett. B 417 (1998) 141; 
              {\it ibid} 427 (1998) 353. 
\bibitem{na93}  R. Narayanan and H. Neuberger, Phys. Rev. Lett. 71 (1993) 3251;
	      Nucl. Phys. B 443 (1995) 305. 
\bibitem{ha98}  P. Hasenfratz, 
              Nucl. Phys. B (Proc. Suppl.) 63A-C (1998) 53; 
                P. Hasenfratz, V. Laliena and F. Niedermayer, 
              Phys. Lett. B 427 (1998) 125. 
\bibitem{gi82}  P.H. Ginsparg and K.G. Wilson, 
              Phys. Rev. D 25 (1982) 2649.

\bibitem{lu98}  M. L\"uscher,
              Phys. Lett. B 428 (1998) 342. 
\bibitem{ch98}  T.-W. Chiu, 
              Phys. Lett. B 445 (1999) 371. 

\bibitem{na97}  R. Narayanan and P. Vranas, 
	      Nucl. Phys. B 506 (1997) 373; 
                R. Narayanan and R.L. Singleton Jr.,
              Nucl. Phys. B (Proc.Suppl.) 63A-C (1998) 555; 
                R.G. Edwards, U.M. Heller and R. Narayanan,
              Nucl. Phys. B 522 (1998) 285; 
              {\it ibid} 535 (1998) 403. 

\bibitem{ch98a}  T.-W. Chiu, 
              Phys. Rev. D 58 (1998) 074511. 

\bibitem{bo85}  M. Bochicchio, L. Maiani, G. Martinelli, G. Rossi and M. Testa, 
	      Nucl. Phys. B 262 (1985) 331. 

\bibitem{br77}  L.S. Brown, R.D. Carlitz and C. Lee, 
              Phys. Rev. D 16 (1977) 417;
                R. Jackiw and C. Rebbi, 
              Phys. Rev. D 16 (1977) 1052.
\bibitem{ke81}  W. Kerler, 
              Phys. Rev. D 23 (1981) 2384; {\it ibid} 24 (1981) 1595;
                E. Seiler and I.O. Stamatescu,
              Phys. Rev. D 25 (1982) 2177; {\it ibid} 26 (1982) 534 (E).
\bibitem{ad98}  D.H. Adams, hep-lat/9812003; 
                H. Suzuki, hep-th/9812019.

\end{thebibliography}
\end{document}